\documentclass[twocolumn,secnumarabic,amssymb,showpacs,superscriptaddress, aps, prl]{revtex4-1}

\usepackage{amsmath} % AMS Math Package
\usepackage{amsthm} % Theorem Formatting
\usepackage{amssymb}	% Math symbols such as \mathbb
\usepackage{graphicx} % Allows for eps images
\usepackage[table]{xcolor}
\usepackage{comment}
\usepackage[export]{adjustbox}
\usepackage{hyperref}% add hypertext capabilities
\hypersetup{
	colorlinks  = true,
	citecolor    = blue
}
\usepackage{blindtext}
\usepackage{enumitem}
\usepackage{lipsum}
\usepackage{verbatim}
\usepackage[normalem]{ulem}

\begin{document}

\title{Transient Hydrodynamic Lattice Cooling by Picosecond Laser Irradiation of Graphite }
\affiliation{Walker Department of Mechanical Engineering, The University of Texas at Austin, Austin, TX 78712, USA}
\affiliation{Department of Mechanical Engineering and Materials Science, University of Pittsburgh, Pittsburgh, PA 15261, USA}
\affiliation{Department of Physics and Astronomy, University of Pittsburgh, Pittsburgh, PA 15261, USA}
\affiliation{Texas Materials Institute, The University of Texas at Austin, Austin, TX 78712, USA}

\author{Jihoon Jeong$^{1}$}
\thanks{These authors contributed equally to this work}
\author{Xun Li$^{2}$}
\thanks{These authors contributed equally to this work}
\author{Sangyeop Lee$^{2,3}$} 
\email{sylee@pitt.edu}
\author{Li Shi$^{1,4}$}
\email{lishi@mail.utexas.edu}
\author{Yaguo Wang$^{1,4}$}
\email{yaguo.wang@austin.utexas.edu}

% Abstract
\begin{abstract}
Recent theories and experiments have suggested hydrodynamic phonon transport features in graphite at unusually high temperatures. Here, we report a pico-second pump-probe thermal reflectance measurement of heat pulse propagation in graphite. The measurement results reveal transient lattice cooling near the adiabatic center of a 15 $\mu$m diameter ring-shape pump beam at temperatures between 80 and 120 K. While such lattice cooling has not been reported in recent diffraction measurements of second sound in graphite, the observation here is consistent with both hydrodynamic phonon transport theory and prior heat pulse measurements of second sound in bulk sodium fluoride.
\end{abstract}

\maketitle
% Introduction
Phonons, the energy quanta of lattice vibration, scatter among themselves via normal (N) and Umklapp (U) processes that conserve and destroy phonon momentum, respectively \cite{peierls1929kinetischen}. The resistivity to heat flow is directly caused by the Umklapp processes and momentum-non-conserving phonon scattering by surface and interior defects.  When these resistive processes are dominant, phonon transport is in the diffusive regime described by Fourier’s law. When the sample size is smaller than the mean free paths of internal phonon-phonon and phonon-defect scatterings, phonon transport is in the ballistic regime. In comparison to the diffusive and ballistic regimes, phonon hydrodynamics describes a peculiar transport regime that is dominated by N-processes. This regime was predicted by the Peierls-Boltzmann transport theory of phonon gas \cite{guyer1966thermal, guyer1966solution} and experimentally observed at temperatures below 20 K in some bulk crystals \cite{ackerman1969second, jackson1970second, mcnelly1970heat, narayanamurti1972observation, mezhov1966measurement} several decades ago. As stated by Ashcroft and Mermin \cite{ashcroft1976solid}, the observation of the hydrodynamic phonon transport is one of the great triumphs of the theory of lattice vibrations. Hydrodynamic phonon transport received a renewed interest in recent years because of first-principles-based theoretical predictions of phonon hydrodynamics in graphitic materials and other two-dimensional (2D) materials near room temperature, which is relevant to technological applications \cite{lee2015hydrodynamic, cepellotti2015phonon, lee2017hydrodynamic, ding2018phonon}.
In the hydrodynamic phonon transport regime, phonon hydrodynamic viscosity \cite{li2018role,ackerman1968temperature} associated with frequent N-scattering increases the phonon-boundary scattering mean free paths and thermal conductance of a rod sample compared to the ballistic transport limit \cite{mezhov1966measurement} due to  momentum-conserving random walk of phonons \cite{gurzhi1964thermal}. In addition, upon pulse heating that is faster than the U-processes but slower than the N-processes, a fluctuating temperature field can propagate in the form of a wave with minimal damping. This thermal wave is referred as second sound, analogous to sound wave propagation in a gas as a pressure fluctuation field due to momentum-conserving intermolecular scattering. The second sound was detected by measuring the temperature evolution at one end of bulk sodium fluoride (NaF) samples at temperatures below 20 K when a heat pulse was applied to the other end \cite{ackerman1969second, jackson1970second, mcnelly1970heat, narayanamurti1972observation}. Recently, the second sound was observed at temperatures as high as about 100 K in a highly oriented pyrolytic graphite (HOPG) sample using a transient grating experiment that detected the diffraction signal from a thermal expansion grating \cite{huberman2019observation}. The measurement result is attributed to the appearance of the peak thermal expansion at the middle in between two heated locations, an unusual feature that can only be caused by second sound propagation. However, the diffraction measurement provided relative thermal expansion difference between different locations without differentiating lattice cooling from lattice heating. In addition, this experiment is different from the direct heat pulse measurement \cite{ackerman1969second, jackson1970second, mcnelly1970heat, narayanamurti1972observation} in the second sound measurements of bulk crystals several decades ago, which revealed detailed thermal fluctuation feature including transient lattice cooling.
In this letter, we report a heat-pulse experiment that is guided by a full solution of the Peierls-Boltzmann transport equation with ab initio inputs to capture the propagation of the second sound in HOPG in space. Below 100 K, we observed a negative displacement of temperature field near the adiabatic center of a 15 $\mu$m diameter heating ring formed by a picosecond laser through a pair of axicon lens, that indicated transient cooling of the lattice. While neither diffusive nor ballistic transport regimes can explain this phenomenon, this lattice cooling feature provides a clear evidence and helps advance detailed understanding of the second sound propagation in graphite.

% figure 1
\begin{figure}[t]
\includegraphics[width=.48\textwidth,center]{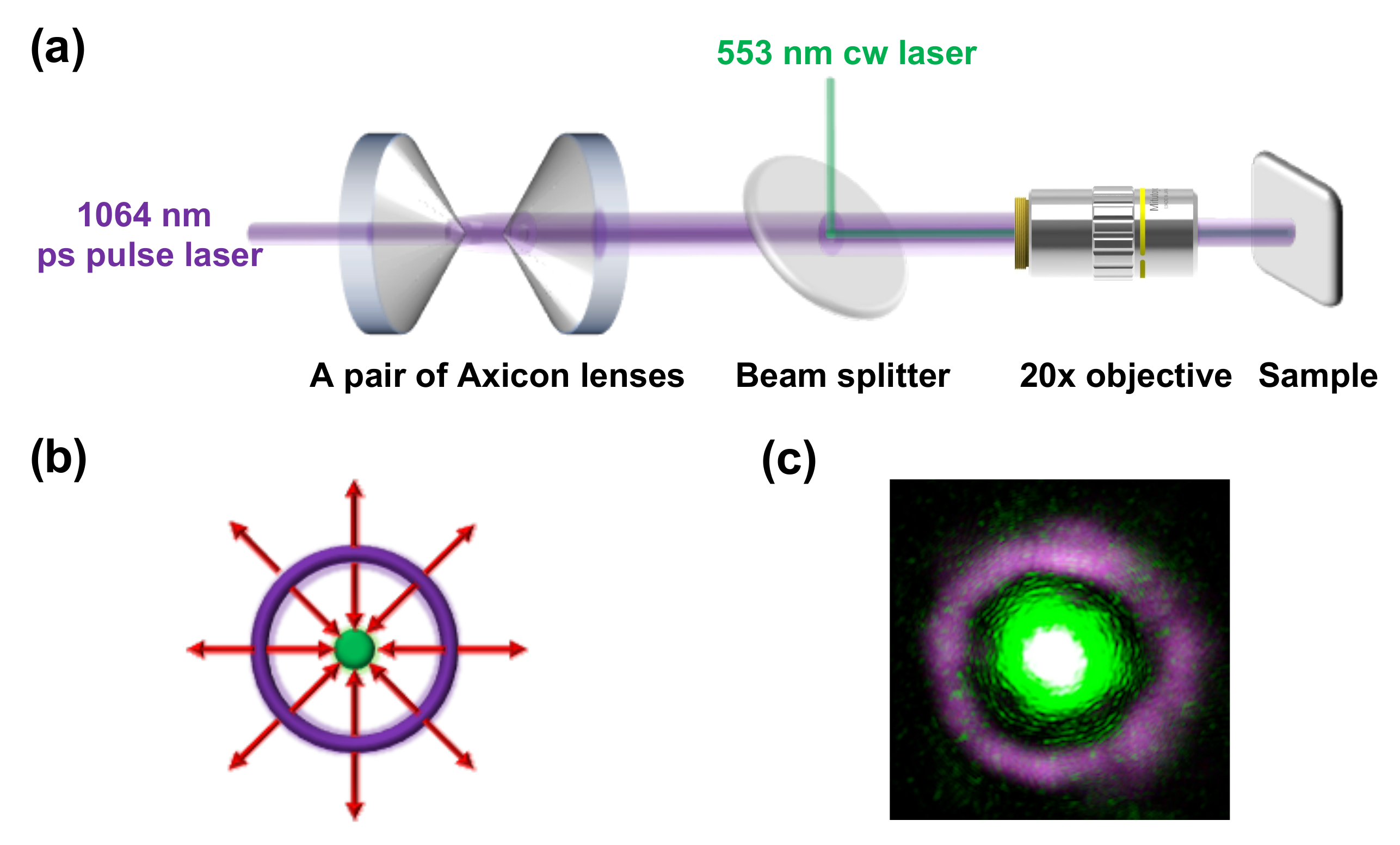}
\caption{Experimental setup of optical heat pulse measurement. (a) Schematic diagram of optical heat pulse measurement setup with a pair of axicon lenses to form a ring-shape pump of 1064 nm picosecond pulse laser. Continuous-wave laser at 553 nm wavelength is focused on the center of the ring-shape pump as a probe. (b) Schematic of the ring-shape pump and probe at the center on the sample surface. Red arrows indicate heat propagating directions. (c) An example image of the laser beam on the sample surface.}
\label{fig:fig1}
\end{figure}

The second sound measurement is implemented with a picosecond transient thermoreflectance (ps-TTR) system \cite{jeong2019picosecond}. The experiment configuration resembles the heat pulse measurement geometry that was employed for the earliest second sound experiments based on a temperature pulse generator and a bolometer \cite{mcnelly1974second, ackerman1969second, jackson1970second}, where a temperature pulse generator and a bolometer were directly attached to each side of the millimeter-size bulk samples, and the second sound propagation is in the millisecond time scale.
A noteworthy difference between the experimental setup for NaF several decades ago and for graphite we report here lies in the spatial and temporal scales. The spatial and temporal scales of second sound propagation are closely related to the mean free paths and rates of N- and U-scattering processes. The second sound propagation can have minimal damping when the wavelength of second sound is in between the mean free paths of N- and U-scattering. Similarly, the frequency of second sound should be in between the rates of N- and U-scattering for the clear observation of the second sound \cite{guyer1966thermal, guyer1966solution}. As the second sound in NaF was observed below 20 K where the phonon scattering rates are relatively lower, detection of temperature with millisecond temporal resolution was sufficient. However, the observation of second sound in graphite at much elevated 100 K would require much smaller temporal and spatial resolutions for the experiment. Therefore, the ps-TTR measurement is designed to provide both sub-nanosecond time resolution and micrometer spatial resolutions.
Our measurement uses a picosecond laser of 1064 nm wavelength as the pump to deliver a heating pulse of 400 ps at 10 kHz repetition rate to the HOPG sample, which is attached with a silver paste to the sample stage of a continuous flow cryostat. A continuous-wave (cw) laser of 553 nm wavelength is employed as a probe, the reflectance of which is detected by the avalanche photodetector (APD) with 500 ps resolution. A pair of axicon lenses is used to form a ring-shape pump beam focused on the sample surface through another 20x objective lens, as shown in Fig. \ref{fig:fig1}(a). The probe beam is delivered to the center of the pump beam on the sample surface to detect the heat pulse propagation, as shown in Fig. \ref{fig:fig1}(b). The 1/$e^2$ diameter of the Gaussian probe beam size is 6 $\mu$m, and the pump ring width is 3 $\mu$m. The average radius of the ring-shape pump beam is calculated as the mean value of inner and outer radii of the ring and is used to estimate the average distance between the pump and probe and the heat pulse propagation distance. Since the in-plane thermal conductivity of graphite is up to 300 times higher than the cross-plane thermal conductivity, the probe beam detects mainly the heat propagation along the basal plane.
% figure 2
\begin{figure}[t]
\includegraphics[width=.36\textwidth,center]{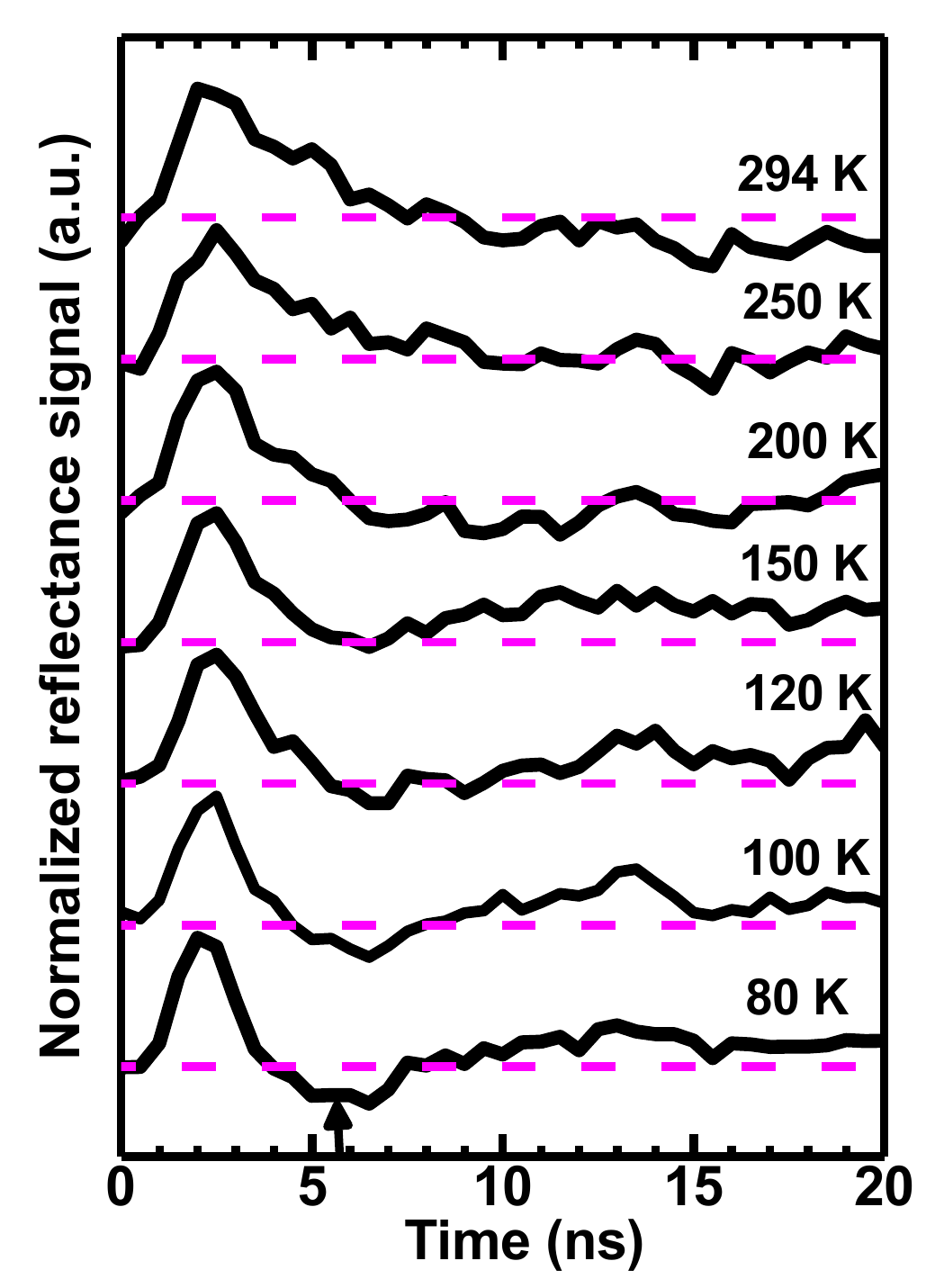}
\caption{Normalized reflectance data measured on the LG-HOPG at different temperatures indicated in the figure. The radius of the ring-shape pump is 15 $\mu$m. The arrow indicates lattice cooling due to second sound propagation. (Note: all the reflectance signals plotted in Fig. 2, Fig. 3 and Fig. 4 are negative values. This is a more convenient way to illustrate the lattice cooling effect because the dR/dT coefficient is negative for graphite.)}
\label{fig:fig2}
\end{figure}
The same heat pulse measurements are conducted on two different commercial HOPG samples with different grain sizes, 10 - 20 $\mu$m for a SPI grade-1 sample denoted as LG-HOPG and shown in Fig. S1 and 30 - 40 nm for a SPI grade-3 sample referred here as SG-HOPG. Fig. \ref{fig:fig2} shows the probe signal measured at the center of the 15 $\mu$m diameter ring-shape pump beam on the LG-HOPG sample at temperature between 300 K and 80 K. Above 150 K, the thermoreflectance signal first arises to the peak and then relaxes to its initial value. This behavior exhibits an ordinary thermal decay that is expected for diffusive phonon transport. As the temperature decreases down to below 120 K, a narrow positive peak appears at about 2 ns delay time, which is followed by a negative peak at around 6 ns. The negative peak becomes increasingly clear when the sample stage temperature decreases further to 80 K. In comparison, the negative temperature displacement is not observed in the SG-HOPG sample at temperatures between 294 and 80 K, as shown in Fig. \ref{fig:fig3}. For three different radii of the ring-shape pump beam, the thermoreflectance signal consistently shows lattice temperature increases and then decays in time in this small-grain size graphite sample.
% figure 3
\begin{figure}[t]
\includegraphics[width=.55\textwidth,center]{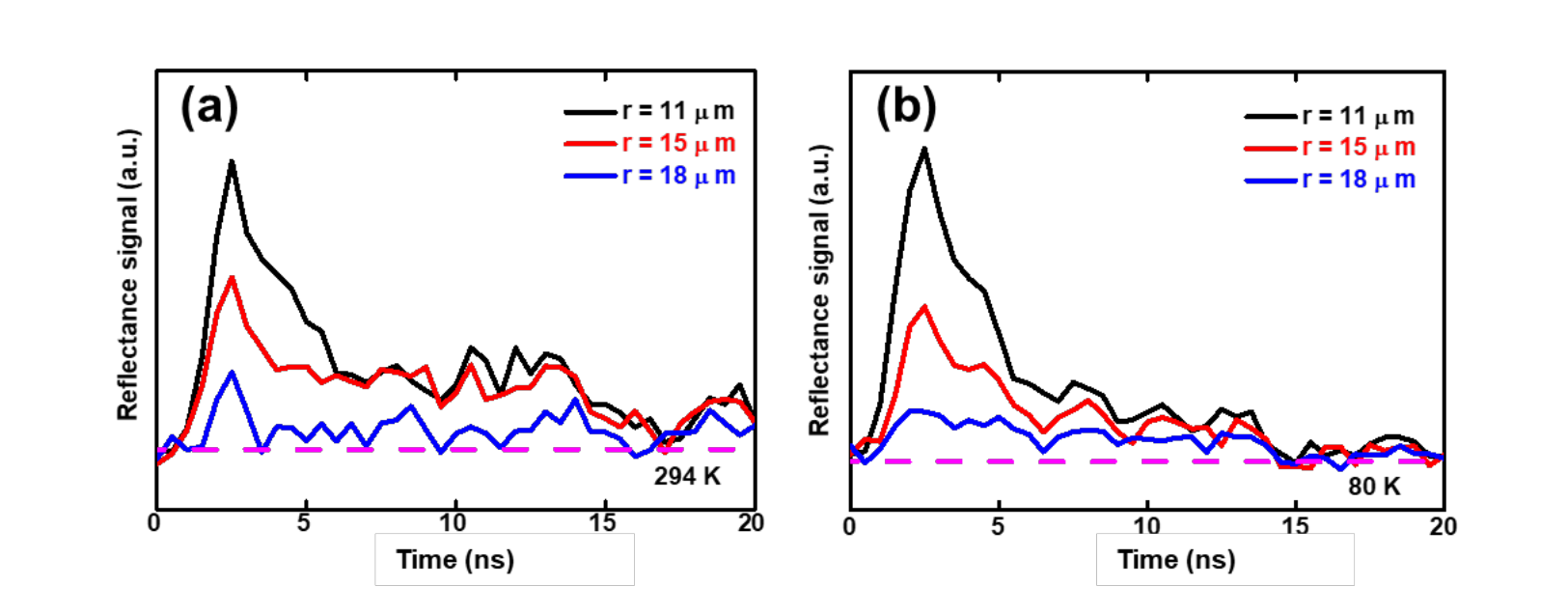}
\caption{Normalized reflectance data measured on the SG-HOPG at (a) 294 K and (b) 80 K. The measurement was conducted at three different radii (r) of the ring-shape pump, including 11, 15, 18 $\mu$m. The horizontal dashed line indicates zero temperature rise.}
\label{fig:fig3}
\end{figure}
When the probe beam is moved to a position about 8.8 $\mu$m away from the adiabatic center of the ring-shape pump beam, the negative temperature rise can still be observed, as shown in Fig. S2.However, when the probe beam is moved to overlap with the ring-shape pump, the temperature rise becomes always positive \cite{seeSM}.We also tested a case without axicon lens, where a Gaussian pump beam and a Gaussian probe beam are offset from each other, the observed temperature rise at the probe beam location is also always positive, as shown in Figs. S3 and S4 \cite{seeSM}. 
The negative temperature displacement, which indicates lattice cooling, cannot occur in either the diffusive or the ballistic transport regime. In the diffusive regime, the pump beam results in a temperature rise that propagates in space and decays monotonically in time. In the ballistic regime, different phonon polarizations gives rise to the propagation of discrete heat pulses that only increases the lattice temperature.  In comparison, in the purely hydrodynamic regime where the total momentum of phonons is conserved, the temperature field can fluctuate between positive and negative displacements without any damping, similar to the pressure fluctuation during sound propagation in gas. The presence of U-scattering results in damping of the fluctuation of the temperature field. Such a damped fluctuation is only observed in the thermoreflectance signal of the large-grain size HOPG sample. Despite the damping, the negative temperature displacement provides an unambiguous evidence of phonon hydrodynamics caused by dominant N-scattering processes in graphite at temperatures between 80 and 120 K. 

To better understand the lattice cooling feature that is observed only near the center of the ring-shape pump beam, we simulated the heat-pulse experiment by solving the Peierls-Boltzmann transport equation in time, real space, and reciprocal space domains with ab initio inputs. To account for the strong N-scattering, the simulation is based on a full matrix of three-phonon scattering instead of commonly used single mode relaxation time approximation \cite{landon2014deviational,li2018role}. The simulation considers the same Gaussian probe beam at the center of the ring-shape heating pulse as in the experiment (Fig. S5). The dimension of the simulated HOPG sample is 20 $\mu$m $\times$ 20 $\mu$m $\times$ 1 $\mu$m \cite{seeSM}.

% figure 4
\begin{figure}[t]
\includegraphics[width=.48\textwidth,center]{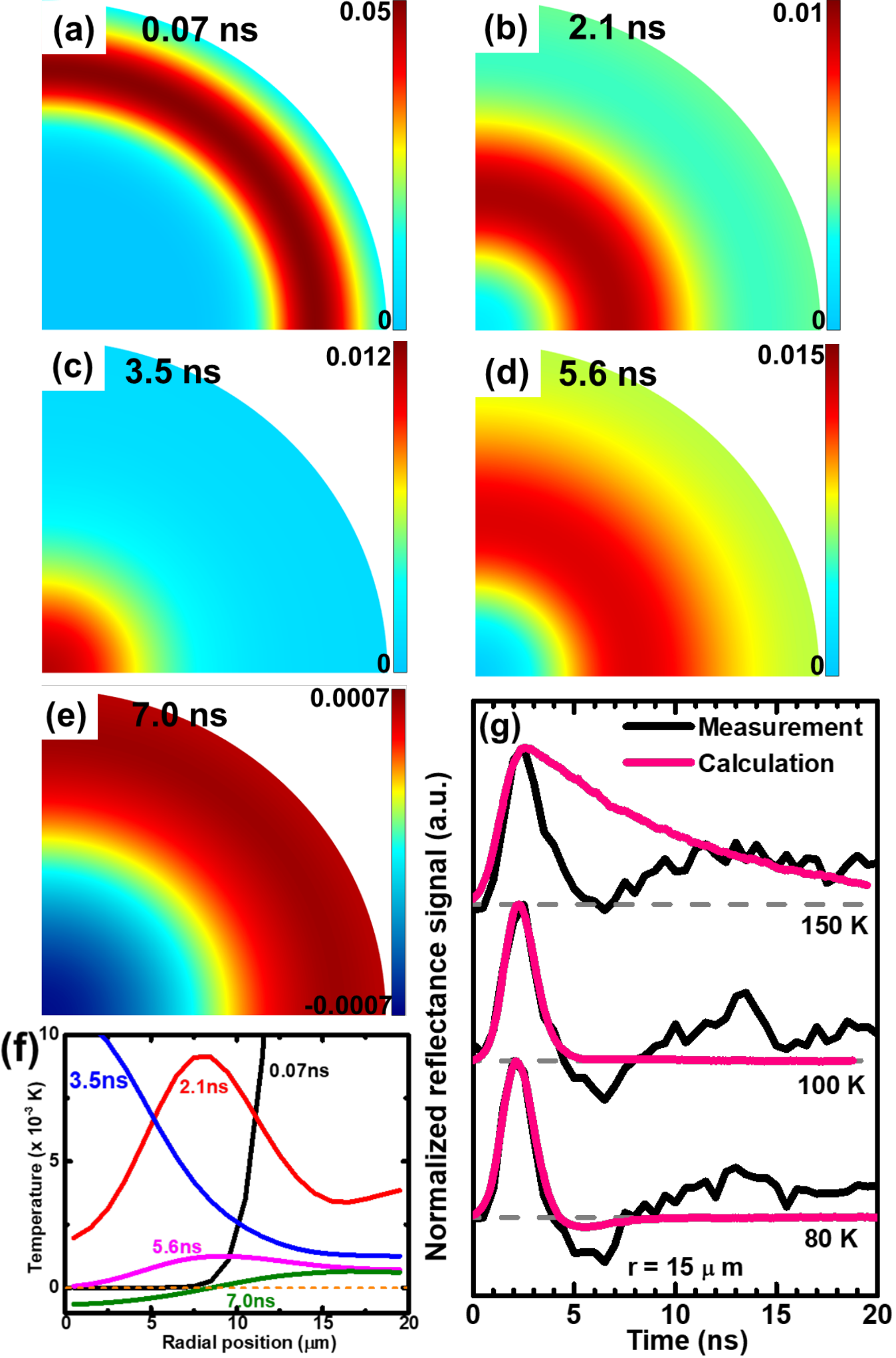}
\caption{(a-e) Snapshots of the calculated transient temperature profiles on the top surface of a graphite sample irradiated by a 15 $\mu$m radiusradius pump beam at an ambient temperature of 80 K. The heat pulse propagates inward in (a-c), then outward in (d), leaving negative temperature rise at the adiabatic center in (e) due to the hydrodynamic thermal fluctuation. (f) Evolution of the temperature profile along the radial direction at each time. (g) Comparison between the experiment and the calculation results for a pump beam radius of 15 $\mu$m at three different temperatures of 80, 100, and 150 K.}
\label{fig:fig4}
\end{figure}

The simulation results in Fig. \ref{fig:fig4} reveal the evolution of temperature profile during the heat-pulse experiment. As shown in Fig. \ref{fig:fig4}(a-b), the applied ring-shape heat pulse propagates toward the center within 2.1 ns following the pulse heating. At 3.5 ns, the heat pulse travelling from the ring-shape heater merges at the center in Fig. \ref{fig:fig4}(c). Subsequently, the heat pulse propagates outward in Fig. \ref{fig:fig4}(d), resulting in negative temperature rise near the center as shown in Fig. \ref{fig:fig4}(e). The evolution of the temperature profile is summarized in Fig. \ref{fig:fig4}(f) to show the clear fluctuation of temperature field at the adiabatic center upon the arrival and departure of the heat pulse. The negative peak predicted is in qualitative agreement with the experimental observations, as shown in Fig. \ref{fig:fig4}(g).

The negative temperature rise observed at the center resembles the results of heat-pulse experiments performed with a bulk NaF sample \cite{jackson1970second}. In the previous report, the temporal temperature profile recorded by a bolometer at the adiabatic end of the bulk sample shows the negative temperature rise upon the arrival of the second sound peak. In our experimental setup with a ring-shape pump, the probe is at the center that is a thermally adiabatic boundary due to the central symmetry. In both the past experiment in NaF \cite{jackson1970second} and our experiment, the negative temperature rise occurred after the second sound pulse is reflected by a thermally adiabatic boundary.
The adiabatic boundary appears to be an important condition for observing the lattice cooling. Indeed, no lattice cooling is observed in both simulation and experiment when the Gaussian pump beam is offset from a Gaussian probe beam \cite{seeSM}. Thus, our experimental setup with the ring-shape pump and probe at the center provides a unique approach to achieve the adiabatic boundary condition at the center. In comparison, an adiabatic boundary condition is also achieved at the middle point between two heating lines in the recently reported diffraction grating measurement of second sounds of graphite \cite{huberman2019observation}. However, the diffraction measurement does not measure the local temperature rise exactly at that middle point. Instead, the diffraction signal is influenced by the spatial distribution of the thermal expansion signal over the entire large probe region without revealing whether lattice cooling could occur at the thermally adiabatic boundary or not.
In summary, our measurement results reveal that, due to hydrodynamic phonon transport, the pulse heating can create lattice cooling near the adiabatic boundary several micrometers away from the heat source in graphite at temperatures between 80 and 120 K. This feature is consistent with both theoretical calculation for graphite and prior second sound measurements of bulk NaF, but cannot be unambiguously revealed from the recent diffraction measurement of second sound in graphite. Therefore, the experimental observation reported here provides detailed insight into the hydrodynamic transport behaviors in graphite.
% Acknowledgement
\begin{acknowledgements}
This work is supported by two collaborative awards (CBET-1705756 and CBET-1707080) from the National Science Foundation. The simulation was performed using the Linux cluster of the Center for Research Computing at the University of Pittsburgh.
\end{acknowledgements}

\bibliographystyle{apsrev4-1}
\bibliography{SS_bibtex}

\end{document}